# Carrier concentration dependence of optical Kerr nonlinearity in indium tin oxide films

Hendry Izaac Elim[1], Wei Ji[1,*], and Furong Zhu[2]

[1]Physics Department, National University of Singapore, 2 Science Drive 3, Singapore 117542

[2]Institute of Materials Research and Engineering, 3 Research Link, Singapore 117602

***Abstract***

Optical Kerr nonlinearity ($n_2$) in *n*-type indium tin oxide (ITO) films coated on glass substrates has been measured using Z-scans with 200-fs laser pulses at wavelengths ranging from 720 to 780 nm. The magnitudes of the measured nonlinearity in the ITO films were found to be dependent on the carrier concentration with a maximum $n_2$-value of 4.1 x $10^{-5}$ $cm^2$/GW at 720-nm wavelength and an electron density of $N_d$ = 5.8 x $10^{20}$ $cm^{-3}$. The Kerr nonlinearity was also observed to be varied with the laser wavelength. By employing a femtosecond time-resolved optical Kerr effect (OKE) technique, the relaxation time of OKE in the ITO films is determined to be ~1 ps. These findings suggest that the Kerr nonlinearity in ITO can be tailored by controlling the carrier concentration, which should be highly desirable in optoelectronic devices for ultrafast all-optical switching.

[*]E-mail address: phyjiwei@nus.edu.sg

# 1 Introduction

Thin films of transparent conducting oxide (TCO) have increasingly attracted attention because they are both electrically conductive and optically transparent in the visible wavelength range. The unique combination of such desirable electrical and optical properties makes TCO films directly relevant to various applications in electronics and optoelectronics [1, 2]. Among many TCOs, indium tin oxide (ITO) is one of the most frequently utilized materials in practical optoelectronic devices. ITO films may be fabricated by many methods, including reactive thermal evaporation deposition, magnetron sputtering, electron beam evaporation, spray pyrolysis, and chemical vapor deposition [3-9]. The optical transparency and electrical conduction mechanism of ITO films have been studied extensively, and important fabrication parameters that control the electrical conductivity of ITO films have been described in detail in Refs [3, 10, 11].

However, the material properties of ITO films have not been fully understood yet, in particular, with regard to the third-order nonlinear-optical properties. Recently, large photo-conductivity in another TCO material, ZnO thin film, has been reported [12], implying that TCOs should possess large optical nonlinearities. In addition, the concentration of free charged carriers (electrons in the conduction band and holes in the valence band) plays a significant role in the electrical conductivity; and as expected, the free-charged-carrier density should influence optical nonlinearities as well, in particular, optical Kerr nonlinearity which is proportional to the light irradiance with the proportionality denoted as $n_2$. Optical Kerr nonlinearity is of direct relevance to all-optical switching devices. Here we report our investigation into optical Kerr nonlinearity in ITO thin films and its dependence on the free-charged-carrier density.

## 2 Experimental arrangements

The nonlinear optical measurements were conducted in the transparent wavelength region by using 200-fs laser pulses at wavelengths ranging from 720 to 780 nm. The ITO films used in the measurements were prepared by a radio frequency magnetron sputtering method described below. An oxidized target with $In_2O_3$ and $SnO_2$ in a weight ratio of 9:1 was employed for the ITO film deposition. The substrate was not heated during and after the film deposition. Sputtering power was kept constant at 100 W. The base pressure in the sputtering system was approximately $2.0 \times 10^{-4}$ Pa. During the film deposition, the argon-hydrogen gas mixture was employed. The argon partial pressure was set as $2.9 \times 10^{-1}$ Pa, and the hydrogen partial pressure was controlled (from $1.1 \times 10^{-3}$ to $4.0 \times 10^{-3}$ Pa) to modulate and optimize the properties of ITO films [13]. Each film was grown on a substrate which consisted of a 1-mm-thick glass coated $SiO_2$ thin film. The film thickness of the TIO films was ranging from 118 to 138 nm, as shown in Table 1. Both Hall effect and Seeback measurements revealed that these films were *n*-type semiconductors and the free-charged-carrier concentrations were varied with the growth conditions. The carrier concentrations of four samples used in the Z-scan measurements discussed below were determined to be $4.0 \times 10^{20}$, $4.4 \times 10^{20}$, $5.4 \times 10^{20}$, and $5.8 \times 10^{20}$ cm$^{-3}$. They are labeled as ITO-2, ITO-3, ITO4, and ITO-5 respectively.

## 3 Results and discussion

Figure 1 displays the transmission spectra of the ITO films, showing that the transparency of the four samples was between 85% and 89% in the spectral region of 700~900 nm. The plots in the inset of Fig. 1 illustrate that the indirect-band-gap energies of the ITO films are located between 3.05 and 3.15 eV, depending on the carrier concentration. As shown in Table 1, the indirect-band-gap energy increases as the free-charged-carrier density rises. It is in agreement with the report by Zhang *et al.* [9]. This dependence is also known as Moss-Burstein shift due to band-gap re-normalization through many-body effects.

To investigate the optical Kerr nonlinearity in the ITO films at room temperature, we employed the Z-scan technique [14] with 200-fs-laser pulses at 1 kHz pulse repetition rate. The laser pulses were generated by a mode-locked Ti:Sapphire laser (Quantronix, IMRA), which seeded a Ti:Sapphire regenerative amplifier (Quantronix, Titan). The laser wavelengths were tunable from 720 to 780 nm (1.72 to 1.59 eV), by feeding the output from the Ti:Sapphire regenerative amplifier to an optical parametric amplifier (TOPAS, Light Conversion, Quantronix). The laser pulses were focused and the sample was Z-scanned across the focal point. By using a set of neutral density filters, the laser irradiances can be varied from a few to 300 GW/cm$^2$ at the focal point.

To eliminate the contribution from glass substrates, we first performed Z-scans on a glass substrate coated with a 200-nm-thick layer of $SiO_2$. Figure 2(a) and (b) display a photo-bleaching signal and positive Kerr nonlinearity ($n_2$), respectively, for the substrate. The photo-beaching in crystalline and amorphous $SiO_2$ has been studied extensively. Recent investigation by Sasajima and Tanimura [15] revealed that the optical bleaching can be attributed to one-center self-trapping holes peaked at 2.16 eV and two-center self-trapping holes for the 2.60-eV band, though the nature of the local distortion that induces the supplemental potential for hole localization is still unclear. Assuming that the absorption coefficient may be expressed as $\alpha_0 + \alpha_2 I$, where $\alpha_0$ is the linear absorption coefficient, $\alpha_2$ the nonlinear absorption coefficient, and $I$ the light irradiance, we employ the Z-scan theory [14] to simulate the open-aperture Z-scan data. The best-fit solid line in Fig. 2(a) infers that $\alpha_2$ is -1 x $10^{-5}$ cm/GW for the substrate.

Figure 2(a) and (b) also illustrate typical open- and close-aperture Z-scans for Sample ITO-5. It is interesting to notice the flat data (within our experimental errors) for the open-aperture Z-scan and the symmetric data for the closed-aperture Z-scan, indicating that absorptive nonlinearity vanishes ($\alpha_2 \sim 0$). We believe that the photo-beaching in the substrate should be cancelled out by the two-photon absorption in the ITO film. The two-photon energy ($2h\nu$ = 3.44 eV for $\lambda$ =720 nm) is greater than the indirect-band-gap energy (3.05 ~ 3.15 eV) of the ITO films. We estimate the two-photon absorption coefficient, $\beta$, for the ITO-5 film by $\beta = -\alpha_2 L_{substrate} / L_{ITO} \sim 0.1$ cm/GW, where $L_{substrate}$ and $L_{ITO}$ are the thickness of the substrate and the ITO film, respectively. Unfortunately, the theoretical value for $\beta$ is unavailable in literature for ITO. However, for indirect-band-gap semiconductor GaP ($E_g^{ind} \approx 2.3$ eV) the theoretical $\beta$-value is reported to be ~ 0.5 cm/GW [16], which is close to our finding for ITO.

Although the resultant nonlinear absorption disappears for the ITO film on the substrate, the resultant nonlinear refraction is enhanced due to the same sign for the Kerr nonlinearity in both the substrate and the ITO film. The difference between the two Z-scans in Fig. 2(b) looks marginal, but the $n_2$ parameter of the ITO-5 film is much greater than the substrate if the 133-nm thickness of the ITO-5 film is compared to the 1-mm thickness of the substrate. By using the Z-scan theory [14] with the total refractive index expressed as $n_0 + n_2 I$, where $n_0$ is the linear refractive index, we numerically compute the close-aperture Z-scans with $n_2$ being an adjustable parameter. The $n_2$ value is extracted by the theoretical curve that best fits to the data. For the substrate, we find that $n_2^{sub}$ is 1.6 x $10^{-13}$ esu (or 4.8 x $10^{-7}$ $cm^2$/GW) at 720 nm, close to ~ 1.1 x $10^{-13}$ esu measured at 1.064 nm by Adair *et al*. [17]. For the ITO film coated on the substrate, the total nonlinear magnitude ($n_2L$) is first extracted from the best-fit Z-scan curve; and then $n_2^{ITO}$ is inferred by $n_2^{ITO} = (n_2L - n_2^{sub}L_{substrate})/L_{ITO}$. Similar experiments and data analyses were carried out for other three wavelengths: 730, 750 and 780 nm. Figure 2(c) displays a typical close-aperture Z-scan for 780 nm while the Z-scan measured at 720 nm is used as a reference. The Kerr nonlinearity in the ITO-5 film was determined to be 4.1 x $10^{-5}$ $cm^2$/GW at 720 nm, 1.2 x $10^{-5}$ $cm^2$/GW at 730 nm, 0.8 x $10^{-5}$ $cm^2$/GW at 750 nm, and 0.6 x $10^{-5}$ $cm^2$/GW at 780 nm. The Z-scans were also conducted at various excitation irradiances. In Fig. 3(a), the $n_2^{ITO}$ values are plotted as a function of the excitation irradiance, which clearly indicates that the observed nonlinearity is of cubic nature. Other three samples (ITO-2, ITO-3, and ITO-4) were investigated in a similar fashion. Their Kerr nonlinearity is summarized in Table 1 while Fig. 3(b) shows the carrier density dependence. Such dependence has been observed in indirect-band-gap semiconductors like Si and Ge [18] and explained by free-electron theory [18]. However, no explicit theoretical calculation has been developed for ITO films yet. Our observation confirms a linear dependence, which has been predicted by the free-electron theory for the low carrier-concentration regime in Si and Ge [18].

To evaluate the relaxation time and to gain an insight of the underlying mechanism for the observed cubic nonlinearities, we conducted a time-resolved optical Kerr effect (OKE) experiment at 780 nm with the use of 200-fs laser pulses from the femtosecond laser system described previously. Figure 4(a) illustrates the transient OKE signals as a function of the delay time for the substrate and Sample ITO-5 at a pump irradiance of 8.5 GW/cm$^2$. The transient signals clearly shows there are two components. By using a two-exponential-component model, the best fits (solid lines in Fig. 4(a)) produce $\tau_1$ = ~ 200 fs and $\tau_2$ = ~ 1 ps. We believe that $\tau_1$ is the autocorrelation of the laser pulses used. The $\tau_2$ component is the recovery time of the excited electrons in the ITO-5 sample. From the OKE signals at zero delay time, the magnitude of the third-order nonlinear susceptibility ( $\chi^{(3)}$ ) of the sample is given by [19]

$$\left|\chi_s^{(3)}\right| \approx \left|\chi_r^{(3)}\right| \left(\frac{F_s}{F_r}\right)^{1/2} \left(\frac{n_0^s}{n_0^r}\right)^2 \left(\frac{L_r}{L_s}\right) \left(\frac{1-R_r}{1-R_s}\right)^{3/2} \left(\frac{\alpha_0^s L_s}{\exp\left(-\frac{1}{2}\alpha_0^s L_s\right)\left[1-\exp\left(-\alpha_0^s L_s\right)\right]}\right) \quad (1)$$

where $F$ is the magnitude of OKE signal, $L$ the interaction length, $R$ the surface reflectance, $\alpha_0$ the linear absorption coefficient, and $n_0$ the linear refractive index. The subscripts $s$ and $r$ denote the ITO-5 sample and the substrate, respectively. As for the substrate (reference sample), the value of $n_2$ is 1.6 x 10$^{-13}$ esu or Re $\chi^{(3)}$ = ~2.5 x 10$^{-14}$ esu; and the value of $\alpha_2$ is -1 x 10$^{-5}$ cm/GW or Im $\chi^{(3)}$ = ~ - 0.5 x 10$^{-14}$ esu, which were measured in the above-discussed Z-scans. Therefore, $\left|\chi_r^{(3)}\right| = \sqrt{\left(\mathrm{Im}\,\chi^{(3)}\right)^2 + \left(\mathrm{Re}\,\chi^{(3)}\right)^2}$ is 2.8 x 10$^{-14}$ esu. Through use of this reference value, the magnitude of $\left|\chi_s^{(3)}\right|$ for the ITO-5 sample is determined to be 1.1 x 10$^{-12}$ esu. This $\chi^{(3)}$ is two orders of magnitude larger than that of the substrate, in agreement with our Z-scan data. On the other hand, Fig. 4(b) displays the time-resolved OKE signals obtained from Sample ITO-5 at the same wavelength with three different pump irradiances. The linear dependence of the OKE signals on the excitation irradiance confirms the nature of third-order nonlinear processes, consistent with our Z-scan studies.

## 4 Conclusions

In conclusion, we have observed the optical Kerr nonlinearity and its carrier concentration dependence in ITO films with femtosecond laser pulses. It should be pointed out that the observed Kerr nonlinearity makes ITO films promising for ultrafast all-optical switching. For an ITO film with an electron density of $N_d$ = 5.8 x $10^{20}$ $cm^{-3}$, the figure of merit [FOM = $n_2/(\beta\lambda)$] is 5.7 at 720 nm, which meets the requirement of FOM > 5 for a Mach-Zehnder-based, all-optical switch [20]. More interestingly, the Kerr nonlinearity in ITO films can be tailored by controlling the carrier concentration. The time-resolved OKE measurements reveal that the recovery time of OKE in the ITO film is ~1 ps.

**Table 1. Electrical properties and nonlinear refractive index of ITO thin films**

| Film properties (Size:1 cm × 1 cm) | Substrate ($SiO_2$) | ITO-2 | ITO-3 | ITO-4 | ITO-5 |
|---|---|---|---|---|---|
| Sheet Resistance ($\Omega$/square) | - | 30.9 | 25.5 | 25.1 | 27.1 |
| Carrier concentration, $N_d$ ($\times 10^{20} cm^{-3}$) | - | 4.0 | 4.4 | 5.4 | 5.8 |
| Mobility ($cm^2$/V-s) | - | 42.9 | 36.8 | 32.5 | 29.1 |
| Thickness (nm) | $1 \times 10^6$ | 118 | 130 | 138 | 133 |
| Conductivity, $\sigma$ ($\times 10^3$ $Scm^{-1}$) | - | 2.91 | 2.61 | 2.81 | 2.70 |
| $E_g^{ind.}$ (eV) | | 3.05 | 3.08 | 3.11 | 3.15 |
| $n_2$ ($\times 10^{-5}$ $cm^2$/GW) at $\lambda$ = 720 nm | 0.048 | 1.6 | 2.1 | 2.6 | 4.1 |
| Re $\chi^{(3)}$ ($\times 10^{-13}$ esu) at $\lambda$ = 720 nm | 0.28 | 15 | 21 | 25 | 40 |

**Figure captions**

**Fig. 1.** Transmission spectra of the ITO thin films coated on glass substrates. From these spectra, the absorption coefficient $\alpha$, is inferred. In the inset, $(\alpha h\nu)^{1/2}$ are plotted as a function of the photon energy, $h\nu$. The thick solid lines represent the straight-line plots of $(\alpha h\nu)^{1/2}$ vs $h\nu$ for determination of the indirect-band-gap energy.

**Fig. 2.**(**a**) Open- and (**b**) closed-aperture Z-scans for the ITO film coated on the substrate recorded at a wavelength of 720 nm and an input irradiance of 256 GW/cm$^2$. For comparison, Z-scans of the glass substrate measured under the same conditions are also displayed. (**c**) Closed-aperture Z-scans of the ITO film coated on the substrate measured at 720 and 780 nm. The symbols are the experimental data. The solid lines are the best-fit curves calculated by the Z-scan theory.

**Fig. 3.** (**a**) The Kerr nonlinearity plotted as a function of the excitation irradiance for the ITO film. (**b**) The Kerr nonlinearity plotted as a function of the free-charged-carrier concentration in the ITO films. The symbols are the experimental data. The solid line in Figure (**a**) is a guide for the eye.

**Fig. 4.** (**a**) Optical Kerr effect (OKE) measurement on the ITO film coated on the substrate. For comparison, OKE measurement on the substrate is also displayed. (**b**) OKE measurements of the ITO film coated on the substrate with three different irradiances.

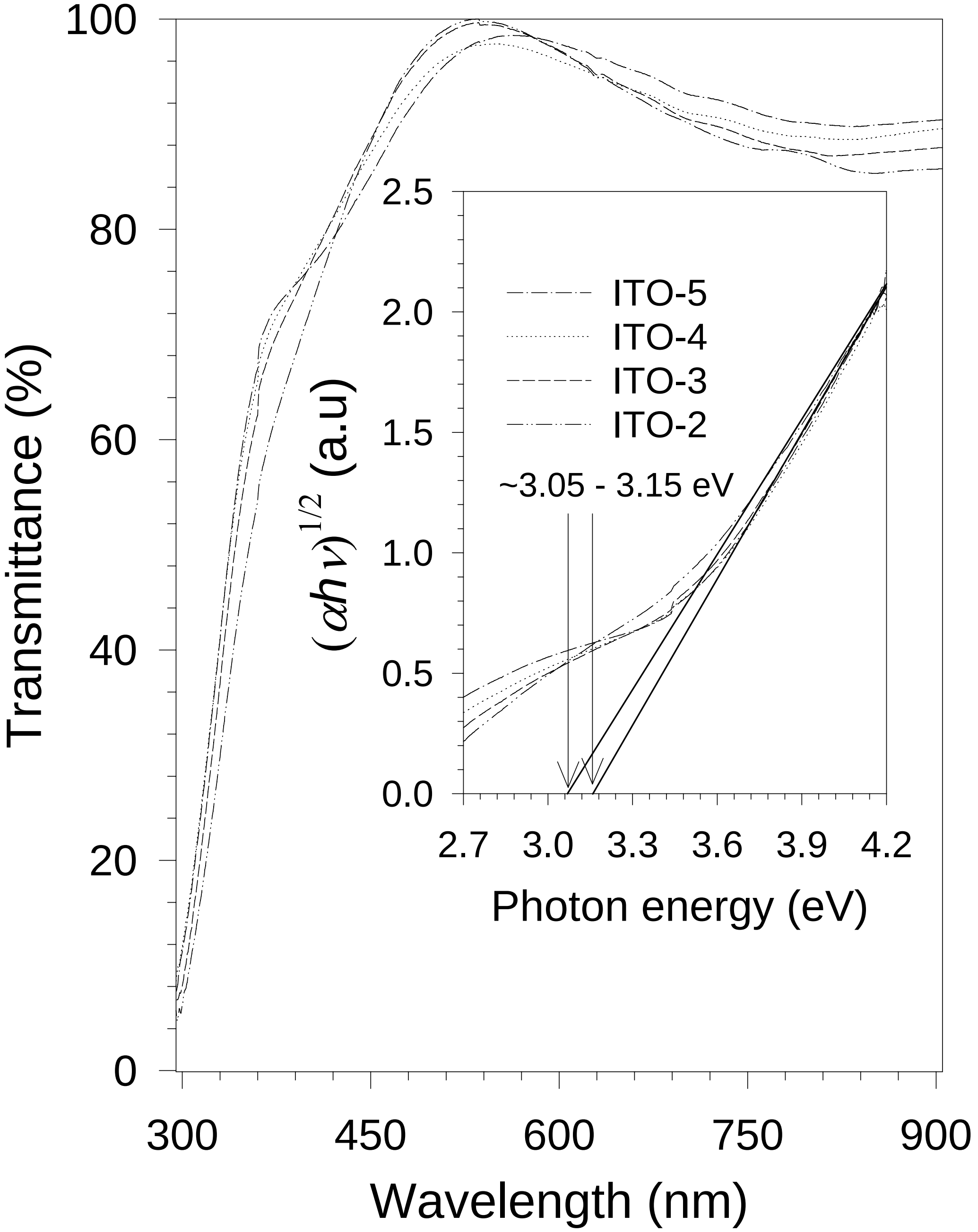

**Figure 1.**

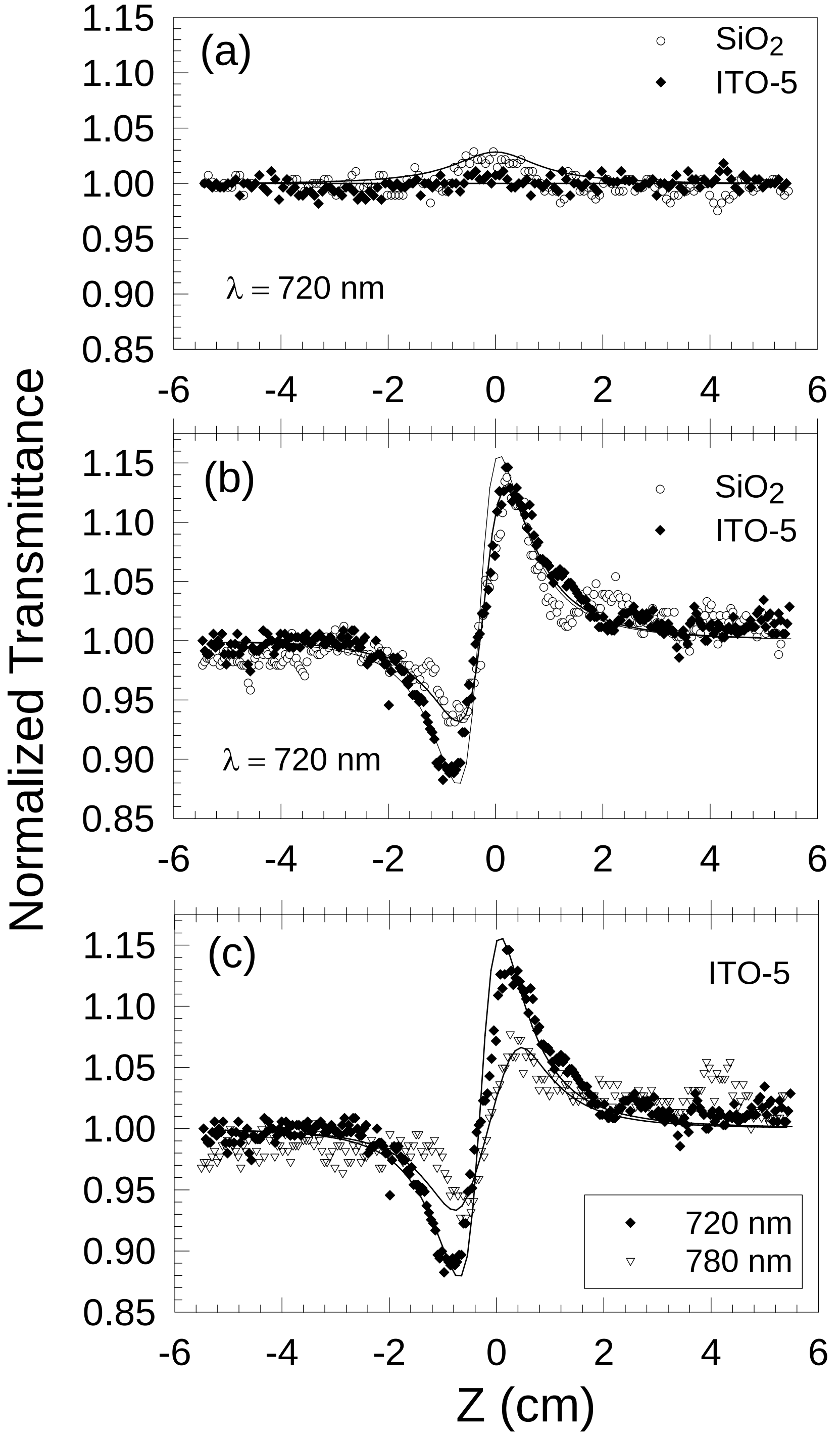

**Figure 2.**

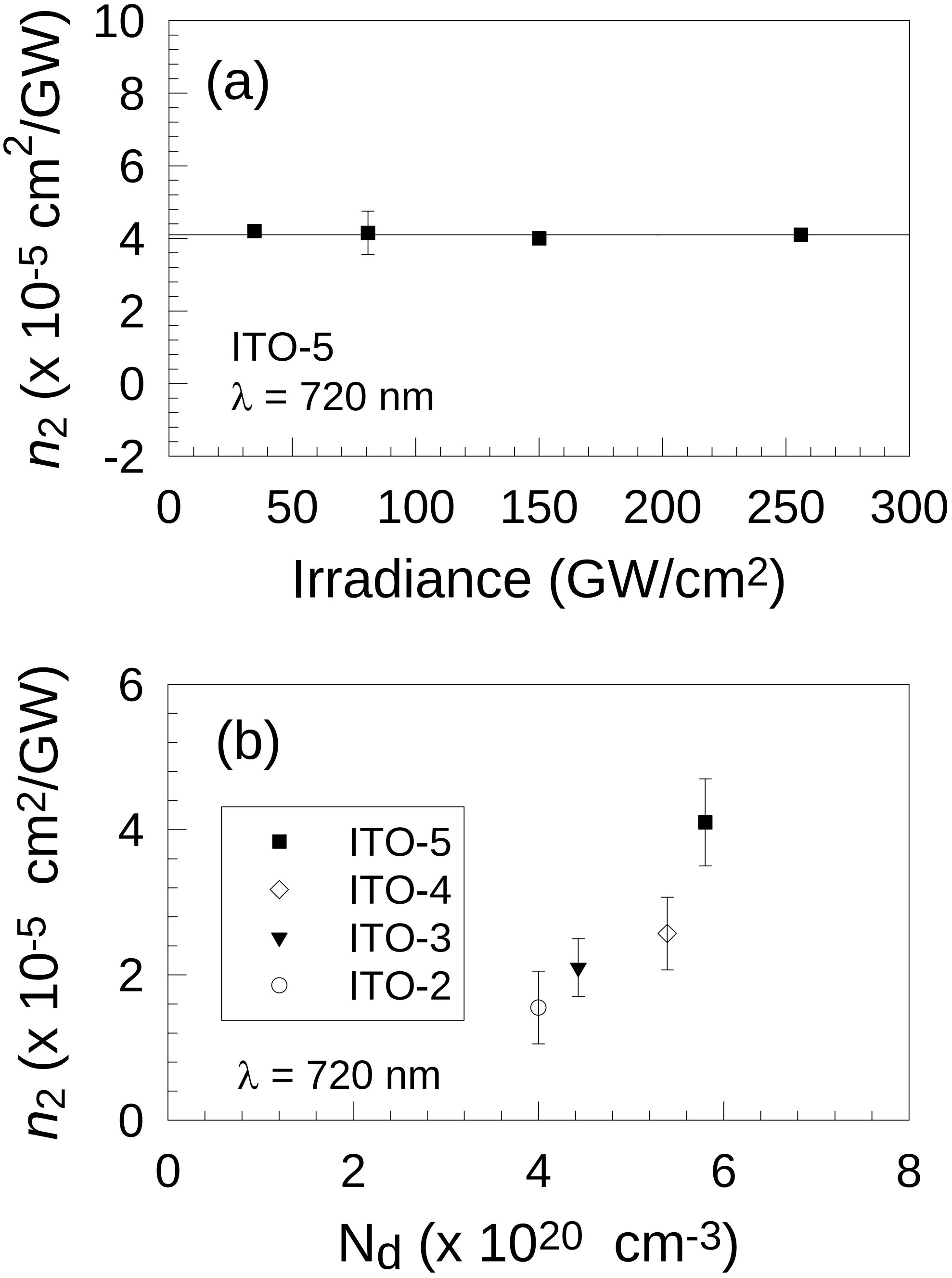

**Figure 3.**

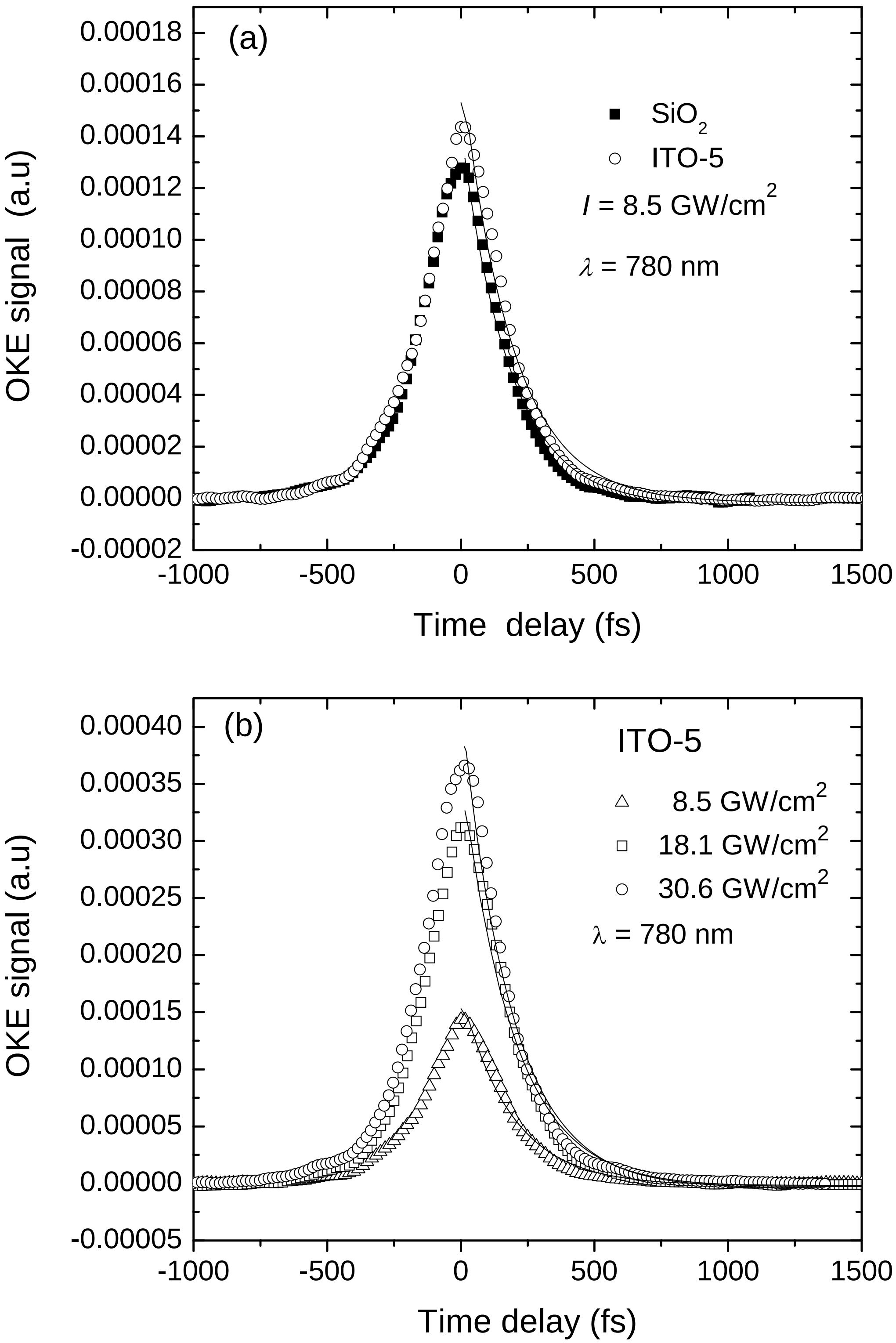

**Figure 4.**